\documentclass{article}
\usepackage{spconf,amsmath,graphicx,tabularx,epstopdf}
\usepackage[table]{xcolor}
\usepackage{cite}

\title{Speech Dereverberation Based on Integrated Deep and Ensemble Learning Algorithm}
\name{Wei-Jen Lee$^{1}$, Syu-Siang Wang$^{1}$, Fei Chen$^{2}$, Xugang Lu$^{3}$, Shao-Yi Chien$^{4}$ and Yu Tsao$^{1}$}
\address{$^{1}$Research Center for Information Technology Innovation, Academia Sinica, Taiwan\\$^{2}$Department of Electrical and Electronic Engineering,\\ Southern University of Science and Technology, China
\\$^{3}$National Institute of Information and Communications Technology, Japan\\$^{4}$Department of Electrical Engineering, National Taiwan University, Taiwan\\}
\begin{document}
\ninept
\maketitle
\begin{abstract}
Reverberation, which is generally caused by sound reflections from walls, ceilings, and floors, can result in severe performance degradation of acoustic applications. Due to a complicated combination of attenuation and time-delay effects, the reverberation property is difficult to characterize, and it remains a challenging task to effectively retrieve the anechoic speech signals from reverberation ones. In the present study, we proposed a novel integrated deep and ensemble learning algorithm (IDEA) for speech dereverberation. The IDEA consists of offline and online phases. In the offline phase, we train multiple dereverberation models, each aiming to precisely dereverb speech signals in a particular acoustic environment; then a unified fusion function is estimated that aims to integrate the information of multiple dereverberation models. In the online phase, an input utterance is first processed by each of the dereverberation models. The outputs of all models are integrated accordingly to generate the final anechoic signal. We evaluated the IDEA on designed acoustic environments, including both matched and mismatched conditions of the training and testing data. Experimental results confirm that the proposed IDEA outperforms single deep-neural-network-based dereverberation model with the same model architecture and training data.
\end{abstract}
\section{Introduction}
\label{sec:intro}
In realistic environments, the perceived speech signal may comprise of the original speech and multiple copies of the attenuated and time-delayed signals \cite{naylor2010speech}. The combination of these signals can cause serious performance degradation of speech-related applications. For example, distant-talking speech significantly degrades the performance of automatic speech recognition (ASR) \cite{feng2014speech,kinoshita2013reverb} and speaker identification \cite{zhao2014robust,sadjadi2011hilbert}. Meanwhile, the adverse effects of reverberation will lower sound quality and intelligibility for both hearing-impaired and normal-hearing listeners \cite{han2015learning,kokkinakis2011channel,roman2013speech}. In the past, various speech dereverberation methods have been developed. The goal of these methods is to extract anechoic speech signals from reverberant ones to enhance the performance of speech-related applications and to improve sound quality and intelligibility simultaneously for listeners in reverberant environments.

Traditional speech dereverberation methods can be roughly divided into three categories \cite{benesty2007springer}. The first category is the source-model-based method, which estimates the clean signal by employing the priori knowledge about time--frequency speech structures \cite{gillespie2001speech,huang2006speech,douglas2003convolutive,li2016speech}. The second category is the homomorphic filtering technique, which adopts a homomorphic transformation to decompose the reverberant signal from the time domain to the cepstral domain, and thus separates the reverberation from the input cepstral coefficients with a simple subtraction operation \cite{bees1991reverberant}. Channel-inversion methods belong to the third category, which considers the reverberation as a convolution of the original sound with the room impulse response (RIR) and thereby performs an inverse filtering to deconvolve the captured signal \cite{nakatani2010speech,kameoka2009robust,mohanan2017speech,wu2006two,kodrasi2014frequency,hikichi2007inverse}. Even though the above three categories of approaches have been shown to provide satisfactory performance, they usually require an accurate estimation of time-varied RIR, which may not always be accessible in practice \cite{xiao2016speech}.

Recently, deep neural network (DNN) models, which show strong regression capabilities, have been used to address the speech dereverberation issue \cite{xiao2016speech,han2014learning}. The main concept here is to use a DNN model to characterize the non-linear spectral mapping from reverberant to anechoic speech in the training stage. In the testing stage, the trained DNN model is used to generate dereverbed utterances given the input reverberant signals. The same concept has been applied to perform denoising and dereverberation simultaneously \cite{han2015learning}. Despite providing notable improvements over traditional algorithms, DNN-based dereverberation methods achieve the optimal performance only in matched training and testing reverberant conditions. To further improve the performance, an environment-aware DNN-based dereverberation system has been proposed, which selects the optimal DNN models online to perform dereverberation \cite{wu2017reverberation}.

Contrary to the idea used in \cite{wu2017reverberation}, the present study extends the previous work on the deep denoise autoencoder (DDAE) in speech enhancement \cite{lu2013speech,lu2014ensemble} and proposes a novel integrated deep and ensemble learning algorithm (IDEA) for speech dereverberation. The IDEA consists of offline and online phases. In the offline phase, multiple DDAE-based dereverberation models are prepared, with each aiming to precisely dereverb speech signals in a particular acoustic environment. Then, a unified fusion model is estimated to integrate the information of the multiple dereverberation models with the aim to estimate clean speech. In the online phase, an input reverberant speech is first processed by all dereverberation models simultaneously, and the outputs are integrated to ultimately generate the anechoic signals. The ensemble learning strategy, which has been proven to be able to improve system performance in speech enhancement \cite{lu2014ensemble} and ASR \cite{tsao2015ensemble,tsao2009ensemble}, is adopted in the task to increase the generalization ability of DDAEs. As will be introduced in the results of experiments, conducted using the Mandarin hearing in noise test (MHINT) \cite{wong2007development}, a DDAE-based dereverberation system achieves the best quality and intelligibility scores when the training and testing conditions are similar (matched condition). However, the performance degrades significantly under mismatched conditions between training and testing. Evaluated results further indicate that the proposed IDEA outperforms the DDAE-based dereverberation system trained in the matched condition and significantly improves speech quality and intelligibility in both matched and mismatched conditions.

The rest of this paper is organized as follows. The spectral-mapping-based speech dereverberation system is reviewed in Section \ref{sec:review}. Then, the proposed IDEA is introduced in Section \ref{sec:IDEL}. Experimental setup and analyses are presented in Section \ref{sec:experiment}. Section \ref{sec:concl} concludes our findings.

\section{Spectral-Mapping-based Speech Dereverberation}
\label{sec:review}
\begin{figure}[!t]
    \centerline{\includegraphics[width=.8\columnwidth]{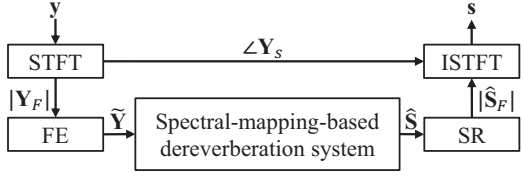}}\vspace{-0.2cm}
    \caption{Block diagram of the spectral-mapping-based speech dereverberation system.}\vspace{-0.2cm}\label{fig:SE}
\end{figure}

In the time domain, the relationship between noisy and clean signals are formulated in Eq. (\ref{eq:assumpt})
\begin{equation}
  \label{eq:assumpt}
  \mathbf{y}=\mathbf{s}\otimes \mathbf{g}+\mathbf{n},
\end{equation}
where $\mathbf{s}$ and $\mathbf{n}$ represent the clean utterance and the additive noise, respectively; ``$\otimes$'' is the operation of convolution; and $\mathbf{g}$ denotes the environmental filter. Fig. \ref{fig:SE} shows the block diagram of the spectral-mapping-based speech dereverberation system, where the goal is to retrieve the anechoic speeches, $\mathbf{x}$, from the reverberant signals, $\mathbf{y}$. As can be seen in Fig. \ref{fig:SE}, $\mathbf{y}$ is first converted to the spectrogram representation $\mathbf{Y}_F$ by carrying out the short time Fourier transform (STFT). Next, a feature extraction (FE) process is conducted to extract the logarithmic power spectrogram (LPS) features $\mathbf{Y}$; then to incorporate the context information, the features $\tilde{\mathbf{Y}}$ are prepared by concatenating the adjacent $M$ static feature frames at the $i$th feature vector $\mathbf{Y}_i$, i.e. $\tilde{\mathbf{Y}}_i=[\mathbf{Y}_{i-M}^\top,\cdots,\mathbf{Y}_{i}^\top,\cdots,\mathbf{Y}_{i+M}^\top]^\top$. The superscript ``$\top$'' denotes the vector transposition. The DNN-based dereverberation system compensates $\tilde{\mathbf{Y}}$ to the estimated LPS $\hat{\mathbf{S}}$ directly, which is further restored to the magnitude spectrum $\vert\hat{\mathbf{S}}_F\vert$ with the spectral restoration (SR) function. Finally, the dereverbed spectrogram $\hat{\mathbf{S}}_F=\vert\hat{\mathbf{S}}_F\vert exp(j\angle \mathbf{Y}_F)$ with an updated magnitude $\vert\hat{\mathbf{S}}_F\vert$ and the original phase $\angle \mathbf{Y}_F$ is converted back to the time domain via inverse STFT (ISTFT) to reconstruct the enhanced time signal $\hat{\mathbf{s}}$.

It is noted that we only consider the reverberant clean signal in Eq. (\ref{eq:assumpt}) and set $\mathbf{n}$ to zero in the present study to focus the dereverberation task.

\section{The Proposed IDEA}
\label{sec:IDEL}
\subsection{Highway-DDAE dereverberation system}\label{sec:hdnntrain}
\begin{figure}[!t]
    \centerline{\includegraphics[width=0.8\columnwidth]{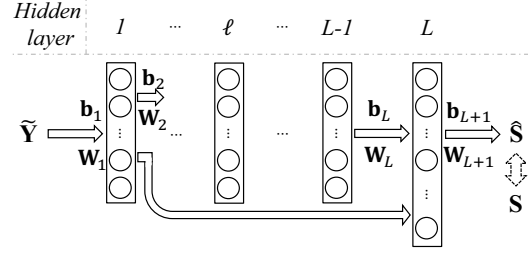}}\vspace{-0.2cm}
    \caption{Flowchart of HDDAE in the offline phase.}\vspace{-0.2cm}\label{fig:DNN}
\end{figure}

In previous studies, traditional fully connected DNNs were used to perform dereverberation \cite{xiao2016speech,han2014learning,wu2017reverberation}. More recently, the highway strategy has been popularly used and shown to provide improved performance \cite{highway}. Our preliminary experiments show that using the highway strategy can improve the speech dereverberation performance in our task. In this section, we first introduce the highway-DDAE (HDDAE). Fig. \ref{fig:DNN} shows the flowchart of the HDDAE for dereverberation in the offline phase.
From the figure, a set of clean--reverb speech pairs ($\mathbf{S}$--$\tilde{\mathbf{Y}}$ pairs) in the LPS domain is prepared first to form the training data, where there are $I$-frame vectors for each of $\mathbf{S}=[\mathbf{S}_1,\cdots,\mathbf{S}_i,\cdots,\mathbf{S}_I]$ and $\tilde{\mathbf{Y}}=[\tilde{\mathbf{Y}}_1,\cdots,\tilde{\mathbf{Y}}_i,\cdots,\tilde{\mathbf{Y}}_I]$. The supervised training procedure is then conducted by placing the clean $\mathbf{S}_i$ and reverb $\tilde{\mathbf{Y}}_i$, respectively, at the output and input sides of the HDDAE model. For the model with $L$ hidden layers, we have:
\begin{equation}\label{eq_forward}
\begin{array}{c}
h_1(\tilde{\mathbf{Y}}_i)=\sigma\{\mathbf{W}_1\tilde{\mathbf{Y}}_i+\mathbf{b}_1\},\\
\vdots\\
h_{\ell}(\tilde{\mathbf{Y}}_i)=\sigma\{\mathbf{W}_{\ell}h_{\ell-1}(\tilde{\mathbf{Y}}_i)+\mathbf{b}_{\ell}\},\\
\vdots\\
h_{L}(\tilde{\mathbf{Y}}_i)=\sigma\{[(\mathbf{W}_{L}h_{L-1}(\tilde{\mathbf{Y}}_i))^\top,(h_{1}(\tilde{\mathbf{Y}}_i))^\top]^\top+\mathbf{b}_{L}\},\\
\hat{\mathbf{S}}_i=\mathbf{W}_{L+1}h_{L}(\tilde{\mathbf{Y}}_i)+\mathbf{b}_{L+1},
\end{array}
\end{equation}
where $\sigma\{\cdot\}$ is a nonlinear mapping function (the ReLu activation function is used in this study). $\mathbf{W}_\ell$ and $\mathbf{b}_\ell$ with $\ell=1,2,\cdots,L+1$ are the weight matrices and bias vectors, respectively. Notably, the output of the $L$th hidden layer $h_{L}(\tilde{\mathbf{Y}}_i)$ cascades $h_{L-1}(\tilde{\mathbf{Y}}_i)$ with $h_{1}(\tilde{\mathbf{Y}}_i)$ (output of the first hidden layer) to possibly address the vanishing gradient problem during the training process (please note that the highway connection may be applied in any two layers; however, the current architecture achieves the best performance in our preliminary experiments). The HDDAE parameter set $\Theta$ consisting of all $\mathbf{W}_\ell$ and $\mathbf{b}_\ell$ are determined accordingly by optimizing the following mean squared error function:
\begin{equation}\label{eq:dnnoptimize}
\Theta^*=\mbox{{\small$\mathop{\arg\min}_{\Theta}(\frac{1}{I}\sum_{i=1}^I\parallel\hat{\mathbf{S}}_i-\mathbf{S}_i\parallel_2^2)$}}.
\end{equation}

\subsection{IDEA for dereverberation}
\begin{figure}[!t]
    \centerline{\includegraphics[width=0.62\columnwidth]{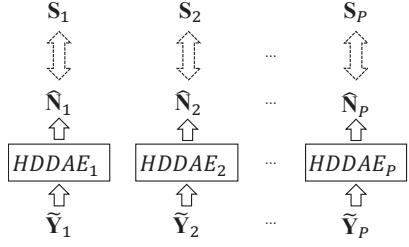}}\vspace{-0.2cm}
    \caption{Flowchart of the EP stage in the offline phase.}\vspace{-0.2cm}\label{fig:basic}
\end{figure}

\begin{figure}[!t]
    \centerline{\includegraphics[width=.97\columnwidth]{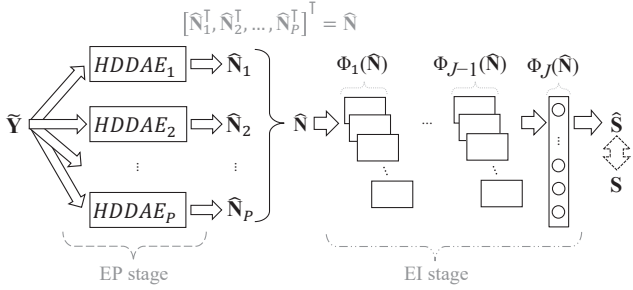}}\vspace{-0.2cm}
    \caption{Flowchart of the IDEA in the offline phase (including the EP and EI stages)}\vspace{-0.2cm}\label{fig:idel}
\end{figure}

In this sub-section, we present the proposed IDEA for speech dereverberation. As mentioned earlier, there are offline and online phases. The offline phase further consists of ensemble preparation (EP) and ensemble integration (EI) stages, which are shown in Figs. \ref{fig:basic} and \ref{fig:idel}, respectively. For the EP stage in Fig. \ref{fig:basic}, there are $1$, $2$, to $P$ reverberant conditions, and thus the reverb data $\tilde{\mathbf{Y}}$ are divided into $P$ subsets, namely, $\tilde{\mathbf{Y}}_{1}$, $\tilde{\mathbf{Y}}_{2}$ to $\tilde{\mathbf{Y}}_{P}$. With these $P$ subsets of training data, together with the corresponding clean training sets, $\mathbf{S}_{1}$, $\mathbf{S}_{2}$ to $\mathbf{S}_{P}$, we have $P$ clean--reverb training sets ($\mathbf{S}_{p}$--$\tilde{\mathbf{Y}}_{p}$ with $p\in\{1,\:\:2,\:\:\cdots,\:\:P\}$). Each training pair is then used to train an HDDAE model. Therefore, the $P$ HDDAE models, $HDDAE_{1}$, $HDDAE_{2}$ to $HDDAE_{P}$, are estimated in the EP stage.


Next, for the EI stage in Fig. \ref{fig:idel}, the input LPS $\tilde{\mathbf{Y}}$ is first processed by the $P$ HDDAE models, as shown in Eq. (\ref{eq:firdnn}).
\begin{equation}\label{eq:firdnn}
\begin{array}{c}
\hat{\mathbf{N}}_{1}=HDDAE_{1}\{\tilde{\mathbf{Y}}\},\\
\hat{\mathbf{N}}_{2}=HDDAE_{2}\{\tilde{\mathbf{Y}}\},\\
\vdots\\
\hat{\mathbf{N}}_{P}=HDDAE_{P}\{\tilde{\mathbf{Y}}\}.
\end{array}
\end{equation}
Then, the outputs of all of these HDDAE models are combined as a new input ($\hat{\mathbf{N}}=[\hat{\mathbf{N}}_{1}^\top,\hat{\mathbf{N}}_{2}^\top,\cdots,\hat{\mathbf{N}}_{P}^\top]^\top$) to train the EI model. In this study, we construct the EI model using a convolutional neural network (CNN) with $J$ hidden layers, as shown in Eq. (\ref{eq:seccnn}), consisting of $J-1$ convolution operations $C_j\{\cdot\}$ at the $i$th sample (frame) vector $\hat{\mathbf{N}}_i$ of the input $\hat{\mathbf{N}}$ and a fully connected hidden layer $F_J\{\cdot\}$.
\begin{equation}\label{eq:seccnn}
\begin{array}{c}
\Phi_1(\hat{\mathbf{N}}_i)=\sigma\{C_1\{\hat{\mathbf{N}}_i\}\},\\
\vdots\\
\Phi_{J-1}(\hat{\mathbf{N}}_i)=\sigma\{C_{J-1}\{\Phi_{J-2}(\hat{\mathbf{N}}_i)\}\},\\
\Phi_{J}(\hat{\mathbf{N}}_i)=\sigma\{F_{J}\{\Phi_{J-1}(\hat{\mathbf{N}}_i)\}\},\\
\hat{\mathbf{S}}_i=F_{J+1}\{\Phi_{J}(\hat{\mathbf{N}}_i)\}.
\end{array}
\end{equation}
The convolution operation applies a set of filters in order to extract $T$ feature maps to obtain local time--frequency structures and to achieve more robust feature representations \cite{fu2016snr}. The provided $\Phi_{J-1}(\hat{\mathbf{N}}_i)$ features at the $(J-1)$th hidden layer are then fed into a fully connected feed-forward network $F_{j}\{\cdot\},\:\:j\in\{J,\:\:J+1\}$, and finally obtain the estimated $\hat{\mathbf{S}}_i$ in the output layer of CNN. Notably, a nonlinear mapping function $\sigma\{\cdot\}$ is applied to modulate the output of each hidden layer. In addition, the parameters $\Lambda$ of the CNN are randomly initialized and then optimized by minimizing the objective function in Eq. (\ref{eq:cnnoptimize}).
\begin{equation}\label{eq:cnnoptimize}
\Lambda^*=\mbox{{\small$\mathop{\arg\min}_{\Lambda}(\frac{1}{I}\sum_{i=1}^I\parallel\hat{\mathbf{S}}_i-\mathbf{S}_i\parallel_2^2)$}}.
\end{equation}

\section{Experiment and Analysis}
\label{sec:experiment}
We evaluated the proposed IDEA using the MHINT sentences \cite{wong2007development} containing 300 utterances pronounced by a native Mandarin male speaker that were recorded in a reverberation-free environment at a sampling rate of 16 kHz. From the database, 250 utterances were selected as the clean training data, and the other 50 utterances were used as the testing data for the speech dereverberation task.

Three distinct reverberant rooms were simulated: room 1 with size $4\times 4\times 4$, room 2 with size $6\times 6\times 4$, and room 3 with size $10\times 10\times 8$, where the unit for all room sizes is meter. The positions of the speakers and receivers were randomly initialized for each room and were fixed for providing RIRs in the considerations of $T_{60}=0.3,\:\:0.4,\:\:0.6,\:\:0.7,\:\:0.9,\mbox{ and }1.0$ (s). For each $T_{60}\in\{0.3,\:\:0.6,\mbox{ and }0.9\}$, three different reverberant environments were provided for deriving RIRs to contaminate the clean training data, and to form the clean--reverb training set accordingly. In addition, one RIR was generated for each of the six $T_{60}$ values to deteriorate all testing utterances and form the testing set. The image model was applied to perform all RIRs by using an RIR generator \cite{habets2006room}. Finally, we prepared $250\times 3(T_{60}\mbox{s})\times 3(\mbox{RIRs})=2250$ and $50\times 6(T_{60}\mbox{s})\times 1(\mbox{RIRs})=300$ reverberant utterances for the training and testing sets, respectively.

In this study, a speech utterance was first windowed to successive frames with the frame size and the shift being 32 ms and 16 ms, respectively. On each frame vector, a 257-dimensional LPS was derived through the STFT and was further extended to $257(2\times 5+1)=2827$ dimensions in terms of $M=5$ mentioned in Section \ref{sec:review} to include the context information as an acoustic feature vector. As a result, the sizes of the input and output layers of the DDAE-based dereverberation system shown in Fig. \ref{fig:SE} were 2827 and 257, respectively. As for the DDAE-based dereverberation system, four types of HDDAE-based architectures were implemented for comparisons: (a) single HDDAE model with three hidden layers ($L=3$ in Eq. (\ref{eq_forward})) trained with the entire training dataset (denoted as ``HDDAE$_{A}(3)$''), (b) single HDDAE model with three hidden layers trained with the dataset composed of one specific $T_{60}$ condition (denoted as ``HDDAE$_{T_{60}}(3)$'' with $T_{60}\in\{0.3,\:\:0.6,\mbox{ and }0.9\}$), (c) single HDDAE model with six hidden layers ($L=6$ in Eq. (\ref{eq_forward})) trained with the entire training dataset (denoted as ``HDDAE$_{A}(6)$'') and (d) the proposed IDEA model (denoted as ``IDEA$_{A}(6)$'') with HDDAE$_{0.3}(3)$, HDDAE$_{0.6}(3)$ and HDDAE$_{0.9}(3)$ in the EP stage, and a CNN model with three hidden layers ($J=3$ in Eq. (\ref{eq:seccnn}); two convolutional layers with each layer containing 32 channels, and a fully-connected layer with 2048 nodes) in the EI stage in Fig. \ref{fig:idel}. Notably, each hidden layer of HDDAEs in (a), (b), (c), and (d) is composed of 2048 nodes.

The speech dereverberation scenarios were evaluated by (a) the quality test in terms of the perceptual evaluation of speech quality (PESQ) \cite{rix2001perceptual}, (b) the perceptual test in terms of short-time objective intelligibility (STOI) \cite{taal2011algorithm}, and (c) the speech distortion index (SDI) test \cite{chen2008fundamentals}. The score ranges of PESQ and STOI are \{-0.5 to 4.5\} and \{0 to 1\}, respectively.
Higher scores of PESQ and STOI denote better sound quality and intelligibility, respectively. On the other hand, the SDI measures the degree of speech distortion, and a lower SDI indicates smaller speech distortions and thus better performance.

\begin{figure}[!t]
    \centerline{\includegraphics[width=.85\columnwidth]{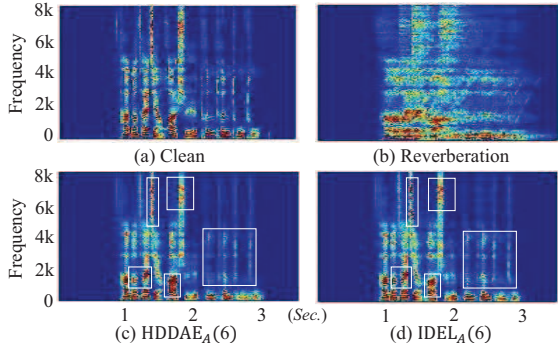}}\vspace{-0.2cm}
    \caption{Spectrum comparison with $T_{60}=1$ s.}\label{fig:spectrum}\vspace{-0.2cm}
\end{figure}

Fig. \ref{fig:spectrum} shows the speech spectrograms corresponding to clean, reverberation at $T_{60}=1.0$ s, processed by HDDAE$_{A}(3)$, and processed by IDEA$_{A}(6)$. From the figure, the spectrogram of the IDEA presents clearer spectral characteristics than those from HDDAE$_{A}(3)$; please note the regions in the white blocks. The harmonic structures for high--frequency components are also clear.

We first list the PESQ scores of HDDAE$_{0.3}(3)$, HDDAE$_{0.6}(3)$ and HDDAE$_{0.9}(3)$ evaluated in either the matched or mismatched testing reverberant conditions in Table \ref{tab:evamatmis}. The results of the baseline (i.e., no dereverberation process was conducted) and HDDAE$_{A}(3)$ are also listed in the table for comparisons. In addition, the averaged PESQ scores (Avg.) for all methods over all testing environments ($T_{60}=0.3,\:\:0.4,\:\:0.6,\:\:0.7,\:\:0.9,\mbox{ and }1.0$) are shown in the last column of the table. In the table, for HDDAE$_{0.3}(3)$, HDDAE$_{0.6}(3)$, and HDDAE$_{0.9}(3)$, the best PESQ score in each of the $T_{60}$ testing conditions is achieved by the HDDAE$_{T_{60}}(3)$ trained on the $T_{60}$ matched condition. In addition, the quality of utterances degrades significantly for those dereverberation systems in the $T_{60}$ mismatched environments, in which the PESQ scores could be even lower than those of baseline (unprocessed input). The observations indicate that the DDAE-based dereverberation system can effectively enhance the speech quality when the property of reverberation is known beforehand, but the performance may degrade dramatically in new environments, where the training and testing conditions are different. Meanwhile, HDDAE$_{A}(3)$ provides the best averaged PESQ score. The result indicates that the model trained on the diverse training set is more robust to varying testing environments.

\begin{table}[!tb]
\begin{center}
\caption{PESQ scores of HDDAE$_{0.3}(3)$, HDDAE$_{0.6}(3)$, HDDAE$_{0.9}(3)$ and HDDAE$_{A}(3)$ testing in either the matched or mismatched reverberant conditions.}\label{tab:evamatmis}
\begin{tabularx}{.9\columnwidth}{>{\centering}m{2.1cm}|ccc|c}
\hline
\hline
\textbf{Testing} $\mathbf{T_{60}}$&$\mathbf{0.3}$&$\mathbf{0.6}$&$\mathbf{0.9}$&Avg.\\
\hline
\hline
\textbf{Reverberation}& 2.0666& 1.5534& 1.1839&1.5661\\
\hline
\textbf{HDDAE$_{0.3}(3)$}& \textbf{2.4830}& 1.4784& 1.0755&1.6373\\
\textbf{HDDAE$_{0.6}(3)$}& 1.6744& 2.2539& 1.2274&1.7072\\
\textbf{HDDAE$_{0.9}(3)$}& 1.3696& 1.6525& 2.1021&1.7217\\
\hline
\textbf{HDDAE$_{A}(3)$}&2.4702&\textbf{2.3064}&\textbf{2.1466}&\textbf{2.2838}\\
\hline
\hline
\end{tabularx}
\end{center}\vspace{-0.4cm}
\end{table}

Table \ref{tab:overall} lists the averaged results of PESQ, STOI, and SDI for unprocessed speech, HDDAE$_{A}(3)$, HDDAE$_{A}(6)$, and IDEA$_{A}(6)$ on all the testing utterances ($T_{60}\in\{0.3,\:\:0.4,\:\:0.6,\:\:0.7,\:\:0.9,\:\:1.0\}$). From the table, we find that all evaluation matrices of DDAE-based approaches outperform those from unprocessed reverberation. These results indicate the effectiveness of the HDDAE-based dereverberation systems. In addition, the better PESQ, STOI and SDI scores of HDDAE$_{A}(3)$ than those from HDDAE$_{A}(6)$ indicate that the additional hidden layers of the HDDAE may not necessarily increase the system performance in the task. On the other hand, IDEA$_{A}(6)$ (also with six hidden layers) yields the highest sound quality and intelligibility and the lowest signal distortion, confirming the effectiveness of the proposed IDEA for the dereverberation task.

\begin{table}[!tb]
\begin{center}
\caption{Averaged results of all testing data for the unprocessed reverberant speech, HDDAE$_{A}(3)$-, HDDAE$_{A}(6)$-, and IDEA$_{A}(6)$-processed utterances.}\label{tab:overall}
\begin{small}
\begin{tabularx}{\columnwidth}{>{\centering}m{0.8cm}|@{\hspace{4pt}}c@{\hspace{4pt}}c@{\hspace{4pt}}c@{\hspace{4pt}}c@{\hspace{4pt}}}
\hline
\hline
\textbf{}&\textbf{Reverberant}&\textbf{HDDAE$_{A}(3)$}&\textbf{HDDAE$_{A}(6)$}&\textbf{IDEA$_{A}(6)$}\\
\hline
\hline
\textbf{PESQ}& 1.5611& 2.2838& 2.2672& \textbf{2.3808}\\
\textbf{STOI}& 0.6692& 0.8598& 0.8527& \textbf{0.8691}\\
\textbf{SDI}& 8.0304& 1.0520& 1.5393& \textbf{0.8916}\\
\hline
\hline
\end{tabularx}
\end{small}
\end{center}\vspace{-0.6cm}
\end{table}

To further analyze the performance of the proposed algorithm, we compare the PESQ scores of IDEA$_{A}(6)$ with those of the HDDAE$_{A}(6)$ in both matched and mismatched testing environments; the results are listed in Tables \ref{tab:finalmatch} and \ref{tab:finalmismatch}, respectively (please note that the testing data in Table \ref{tab:finalmismatch} cover $T_{60}=\{0.4,\:\:0.7,\:\:1.0\}$, which were not seen in the training data). From these tables, we observe that PESQ scores obtained by IDEA$_{A}(6)$ and HDDAE$_{A}(6)$ consistently decrease with increasing $T_{60}$, revealing that the dereverberation performance is negatively correlated with the $T_{60}$ value. In addition, IDEA$_{A}(6)$ outperforms HDDAE$_{A}(6)$ in all testing $T_{60}$s, confirming that the ensemble modeling can achieve better results than those from a single model, where the training data and the number of layers are the same for these two models.

\begin{table}[!tb]
\begin{center}
\caption{PESQ scores of HDDAE$_{A}(6)$ and IDEA$_{A}(6)$ evaluated in the matched testing conditions}\label{tab:finalmatch}
\begin{tabularx}{0.8\columnwidth}{>{\centering}m{2.2cm}|ccc}
\hline
\hline
\textbf{Testing} $\mathbf{T_{60}}$&$\mathbf{0.3}$&$\mathbf{0.6}$&$\mathbf{0.9}$\\
\hline
\hline
\textbf{HDDAE$_{A}(6)$}& 2.4349& 2.2990& 2.1408\\
\textbf{IDEA$_{A}(6)$}& \textbf{2.5669}& \textbf{2.4249}& \textbf{2.2479}\\
\hline
\hline
\end{tabularx}
\end{center}\vspace{-0.6cm}
\end{table}
\begin{table}[!tb]
\begin{center}
\caption{PESQ scores of HDDAE$_{A}(6)$ and IDEA$_{A}(6)$ evaluated in the mismatched testing conditions}\label{tab:finalmismatch}
\begin{tabularx}{0.8\columnwidth}{>{\centering}m{2.2cm}|ccc}
\hline
\hline
\textbf{Testing} $\mathbf{T_{60}}$&$\mathbf{0.4}$&$\mathbf{0.7}$&$\mathbf{1.0}$\\
\hline
\hline
\textbf{HDDAE$_{A}(6)$}& 2.3575& 2.2309& 2.1399\\
\textbf{IDEA$_{A}(6)$}& \textbf{2.4676}& \textbf{2.3323}& \textbf{2.2452}\\
\hline
\hline
\end{tabularx}
\end{center}\vspace{-0.4cm}
\end{table}
\vspace{-0.2cm}
\section{Conclusion}
\label{sec:concl}
From the experimental results, we first noted that the single-HDDAE-based systems could achieve good dereverberation performance in matched conditions, but the performance degraded significantly when the systems were tested in mismatched conditions, showing that the HDDAE models trained to address specific reverberation conditions may have limited generalization capabilities. In addition, the model HDDAE$_{A}$, which was trained using all the training data, outperformed individual HDDAE models in terms of PESQ scores over all testing environments. Moreover, when compared to the model HDDAE$_{A}$, the model IDEA$_{A}$ provided better results, confirming that by collecting information from multiple environments to train matched HDDAE models and then integrating the information from the outputs of these models, diverse reverberation conditions can be covered and high dereverberation performance achieved.

\section{Acknowledge}
This research was supported in part by the Ministry of Science and Technology of Taiwan (MOST 107-2633-E-002-001), National Taiwan University, Intel Corporation, and Delta Electronics.

\bibliographystyle{ieeetr}
\bibliography{reference}

\end{document}